\documentclass[a4paper]{jpconf}
\usepackage{graphicx}
\usepackage{array}

\begin{document}
\title{A scaffolded curriculum to foster experimental skills acquisition in an introductory physics lab course}

\author{M. Alemani$^1$}

\address{$^1$ Physics and Astronomy Institute, University of Potsdam, Potsdam 14476, Germany\\ 
ORCID: 0000-0003-2004-8565}

\ead{alemani@uni-potsdam.de}

\begin{abstract}
This article outlines a comprehensive redesign of an introductory physics laboratory course aimed at promoting active student engagement, enhancing scientific experimental skills and the development of expert-like attitudes. The redesign process involved redefining specific and measurable learning goals and the development of tailored activities and assessment rubrics to reach and assess those goals. In order to implement course scaffolding, we made significant modifications to the previously existing course format, space, physics topics contexts, and assessment. Instructors' roles evolved from imparting knowledge to promoting student decision-making and fostering a collaborative learning environment. To guide the laboratory transformation we used the results of a research-based survey.

This work offers practical insights for educators seeking to redesign laboratory courses, emphasizing authentic scientific practices and student-driven learning experiences in physics education.

\end{abstract}

\section{Introduction}\label{Introduction}

Teaching methodologies that actively involve students in the collaborative construction of knowledge have demonstrated greater efficacy compared to passive information reception. A comprehensive review of the literature on this subject is available in reference \cite{Meltzer}. Physics laboratory courses (PLCs) could be considered inherently active learning environments, as students are active when working on lab activities. Recent research has however raised doubts regarding the efficacy of traditional prescriptive PLCs in enhancing physics content knowledge \cite{Wieman}, developing a deeper understanding of measurement uncertainties \cite{Volkwyn}, and fostering positive attitudes toward experimental physics \cite{Wilcox1}.

Reflecting on these results, one could argue that \textit{hands-on }activities do not necessarily mean \textit{minds-on} activities 
\cite{Redish} since just ‘interpreting’ a prescriptive lab manual is not engaging students with higher cognitive levels, which eventually involve (i) the creation of new outputs by students beyond the information received by the instructors as well as (ii) collaborative aspects of learning \cite{ICAP}. The comparison of the cognitive tasks used by professional experimental physicists with those performed by students in a traditional confirmatory introductory laboratory course shows that the majority of the cognitive tasks are already done by instructors \cite{WiemanCognitive}. 

A recent historical review on instruction in PLCs in the United States highlights that there have been recurrent and consistent criticisms of traditional prescriptive instruction in PLCs in the past 200 years \cite{HistReview}. 
However, only in recent years a shift in focus from laboratory course structures to their learning goals had happened.

Consequently, novel active learning pedagogies have been proposed and their effectiveness has been systematically studied using rigorous evidence-based methods \cite{Wieman, Volkwyn, Etkina, Etkina2, Zwickl}. 

Nevertheless, despite studies highlighting the deficiencies of the traditional laboratory courses, traditional prescriptive and confirmatory labs endure in numerous physics programs.
On the other hand, the working group of the GIREP community with focus on laboratory instruction recently wrote a position paper \cite{GIREP_PosPaper_Labs} stating that the main goal of lab courses should be to (i) Help students to learn how to think like a physicist/scientist, (ii) help students to learn how to plan, design and conduct experiments and experimental procedures, and (iii) help students to learn how to collect, analyze, present data and report the findings.

In light of the literature discussed above and aligned with the reflections emphasized by the GIREP work group on laboratory courses, we have undertaken a redesign of our PLC for physics major students at the University of Potsdam (UP). Our redesign entails a transition from a conventional 'teacher-focused' and 'concept-based' approach to a more 'student-centered' and 'skill-based' paradigm. This shift aims to promote cognitive development and foster collaborative engagement among students during learning activities as well as students understanding of the nature of experimental physics.
While the assessment of our course transformation was discussed in detail elsewhere \cite{Teichmann, Teichmann2, Alemani_course}, we want to focus here on the discussion of the development of a scaffolded curriculum to foster experimental skills and expert-like attitudes acquisition.

\section{Existing Course Description}\label{CourseDescrip}
The introductory laboratory course here described addresses physics major students. It is composed of a sequence of four courses which take place during the first four semesters of the physics curriculum. Typically there are between 40 and 60 students enrolled in each instance of the course. Each part of the course is obligatory and except the last part of the sequence, it is not graded. 

Before the transformation students were doing overall 30 different experiments, one per lab appointment. Students were working on different physics topics encountered in the physics lecture for the allocated time of about three hours. All experiments had the same structure: Students were given a setup already built and previously optimized, received a detailed list of instructions to interpret and follow for the data taking and the data analysis with the aim of reproducing well known results from theory. 
They were confronted with many different measurement and data analysis techniques but were not asked or allowed to engage in any of the steps involved in designing an experiment or choosing a data analysis strategy.
There was also no space for iteration of experimental steps, the experimental process was linear. The sessions ended with the lab report homework either concluding that the theory was confirmed by the experiment or in case of disagreement by proposing reasons for it. The success of the experiment was biased towards the end result. Experimental problems were usually experienced by students as a source of frustration. 

As evidenced by existing literature, our experience revealed insufficient levels of student engagement \cite{WiemanCognitive, Holmes}, identified a gap in the development of crucial experimental skills \cite{Volkwyn}, and expert-like attitudes \cite{Wilcox1}. To address these shortcomings, we initiated a comprehensive transformation of our course. In the subsequent sections, we provide a detailed account of the steps involved in this transformative process.

\section{Model for the course transformation}\label{Process}
The process of redesigning a laboratory course is inherently complex, involving a multitude of steps and considerations. 
To effectively navigate this complexity, we employed a backward design methodology, as proposed by Zwickl et al. \cite{Zwickl4}. 

\begin{figure}
\begin{center}
\includegraphics[width=0.6\textwidth]{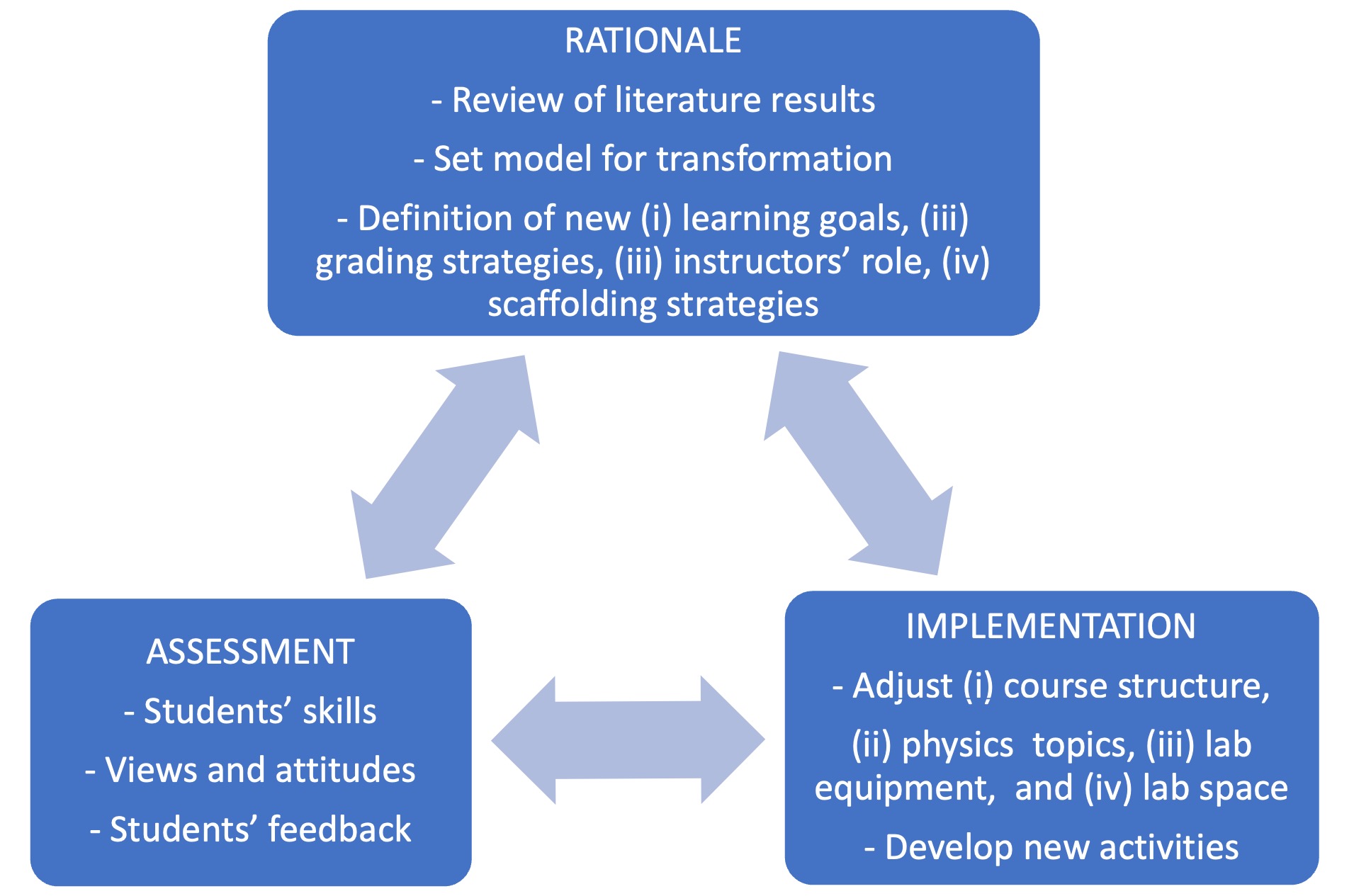}
\end{center}
\caption{\label{fig0} Visualization of the lab redesign process with its steps and components.}
\end{figure}

%beginning of new text
Figure \ref{fig0} provides a simple visualization of the process we undertook. Our approach began with delineating our rationale, which involved reviewing the literature on labs, setting a transformation strategy, and defining new learning goals, grading methods, and course scaffolding. This was followed by implementation, which included adjusting the course structure, physics topics, lab equipment, redesigning the lab space, and developing tailored activities to achieve the learning goals. The final step involved assessing students' learning.
It is important to note that these three aspects of the lab course redesign are interconnected (as shown by the arrows in figure \ref{fig0}) and were iteratively revised based on the assessment results. 
The following sections will detail the different aspects related to the first two steps of the lab redesign, while the assessment process will be briefly outlined.
% end new text

% Old text
%This approach entails delineating clear learning goals for students as a primary step, followed by the development of tailored activities aimed at achieving these goals, and subsequently assessing the attainment of these goals based on student learning outcomes. In the following, we describe the first two steps of the process, i.e. the definition of the learning goals and the transformations to the curriculum, while we only briefly outline the assessment process.

%beginning new figure

\section{Course learning goals, objectives and outcomes}\label{LearningGoals}
Our goal is to foster students' acquisition of scientific experimental skills. To classify those skills we used the framework developed by Zwickl et al. \cite{Zwickl4}  and divided them into the categories \textit{design}, \textit{modeling}, \textit{technical-computational}\footnote{Note that we add the computational aspect to the technical skills category of Zwickl et al. \cite{Zwickl4}} and \textit{communication}. 
As a further goal we seek to promote the development of experts-like views and attitudes towards experimental physics. We do this by offering students authentic laboratory experiences in which students engage in scientific practices. In fact, we consider this second aspect crucial for students’ achievement of the first learning goal.

Based on these general goals, we defined the following course-wide objectives phrased here as learning outcomes:
\textit{By the end of the four courses of the introductory laboratory sequence, students should be able to:}
\begin{itemize}
    \item Design measurement setups, methods, procedures and measurement strategies 
    \item Use and refine models of physical and measurement systems to make sense of experimental results both qualitatively and quantitatively based on evidence 
    \item Use available instrumentation and programming environments to collect and analyze data or manipulate the real world
    \item Communicate clearly processes and outcomes of an experiment supporting arguments by evidence and using formats utilised by professionals
\end{itemize}

\begin{figure}
\begin{center}
\includegraphics[width=0.9\textwidth]{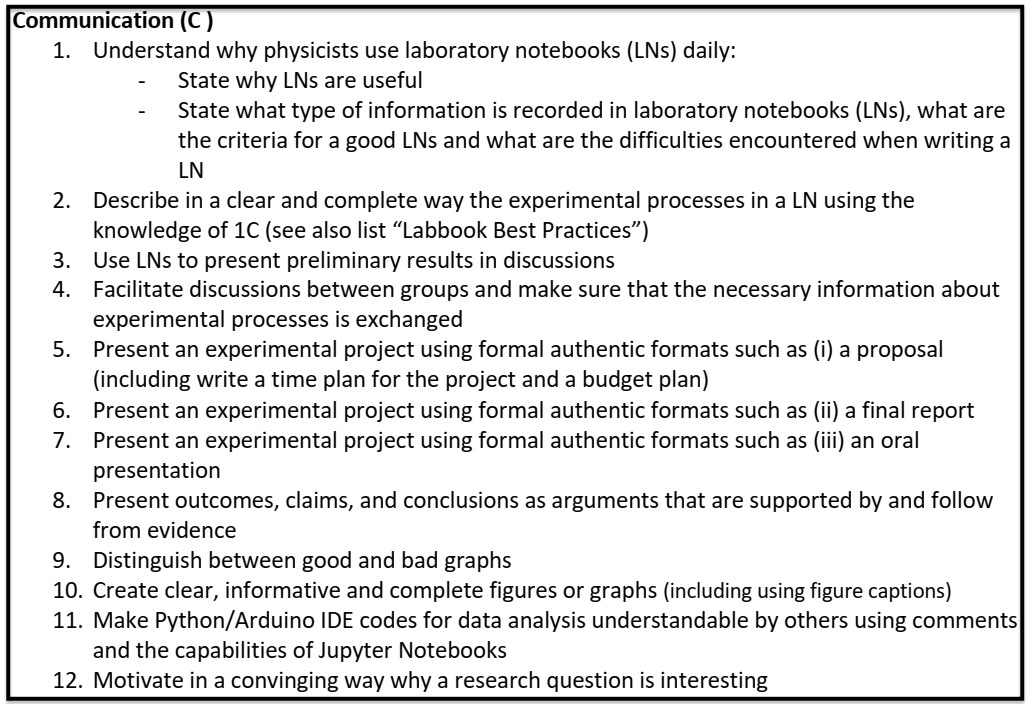}
\end{center}
\caption{\label{fig1} Specific learning outcomes for the laboratory course sequence in the skills category \textbf{Communication}}
\end{figure}

Finally for each category of skills, we have meticulously outlined a comprehensive set of specific and measurable learning outcomes (LOs). Each LO is accompanied by designated activities designed to facilitate student practice and assessment. These LOs represent finely detailed educational objectives. Several of these LOs could be grouped into broader categories to streamline the learning process.
We intentionally defined very specific LOs to be able to scaffold practice activities.
The list of our specific LOs for the category \textit{communication} is shown as an example in figure \ref{fig1}. The complete list including the categories \textit{design} and \textit{modeling} and \textit{technical and computational} can be found in \cite{OSF}.
To formulate these LO we made use of the revised Bloom taxonomy \cite{BloomRev}. This way, we checked that all levels of the cognitive domain are involved.
Note that these LOs are the results of various iterations. For example we validated those LO by re-examining our activities and grading rubrics and monitoring how often each LOs was practiced by students. After such a check, one might find out that some of the defined LOs are not actually requested for the successful achievement of the lab activities.

\section{Modifications imparted to the existing laboratory course}\label{curriculum}

To allow the achievement of the newly established learning goals, we made use of scaffolding techniques. Consequently significant modifications have been implemented in the existing laboratory course. These alterations encompass various facets, such as the redesign of the laboratory space and the course structure. In the subsequent sections, we provide an overview of each of these modifications.

\subsection{Scaffolding}\label{Scaffolding}
Conducting experiments necessitates the utilization of a diverse array of skills. We counted up to about 30-40 different of our 'fine-grained' skills per experimental activity. To ensure that students engage in targeted practice to master each of these skills, we structured our curriculum and activities using scaffolding techniques.

Rather than overburdening students with a multitude of new skills from the beginning, we strategically staggered the introduction of new LOs in each lab activity (up to 11), thereby ensuring that students encounter a manageable number of LOs during each session. In the new curriculum, students initially practice fundamental skills during specialized active learning tutorials conducted in the first and second semester. These focus specifically on essential aspects of experimentation like estimation of measurement uncertainties, (visual) data representation and analysis and authentic use of laboratory notebooks \cite{Stanley, Stanley2, Hoehn, Alemani_Labbook1, Alemani_Labbook2}.
In other tutorials we introduce computational aspects of data handling and analysis in Python using Jupyter Notebook \cite{Tufino}. In a tutorial at the beginning of the second semester, students are introduced to the 'Modeling Framework for Experimental Physics' developed by H. Lewandowski and colleagues \cite{Zwickl, Zwickl3}. This framework was developed to guide students during experiments. It clarifies the importance of modeling the experimental measurement system in addition to the physical system to be studied, and highlights the iterative aspect in modeling.

Furthermore, our curriculum is sequenced such that more complex cognitive tasks are introduced after students have gained proficiency in foundational skills. For instance, students first learn to comprehend and apply data analysis methods before progressing to the development of their own data analysis strategies within experiments. Similarly, students initially practice documenting experiments using instructional videos before transitioning to documenting their own experiments independently. Note that, the difference to traditional concepts is the focus on the LOs, not mixing them with all other goals in pre-defined experimental work.
For the lab activities, this type of scaffolding is realized also by gradually increasing the complexity of experimental demands. However, it is important to note that one should not confuse the complexity of experimental setups with the level of the cognitive tasks that students engage with.

\begin{figure}
\begin{center}
\includegraphics[width=1\textwidth]{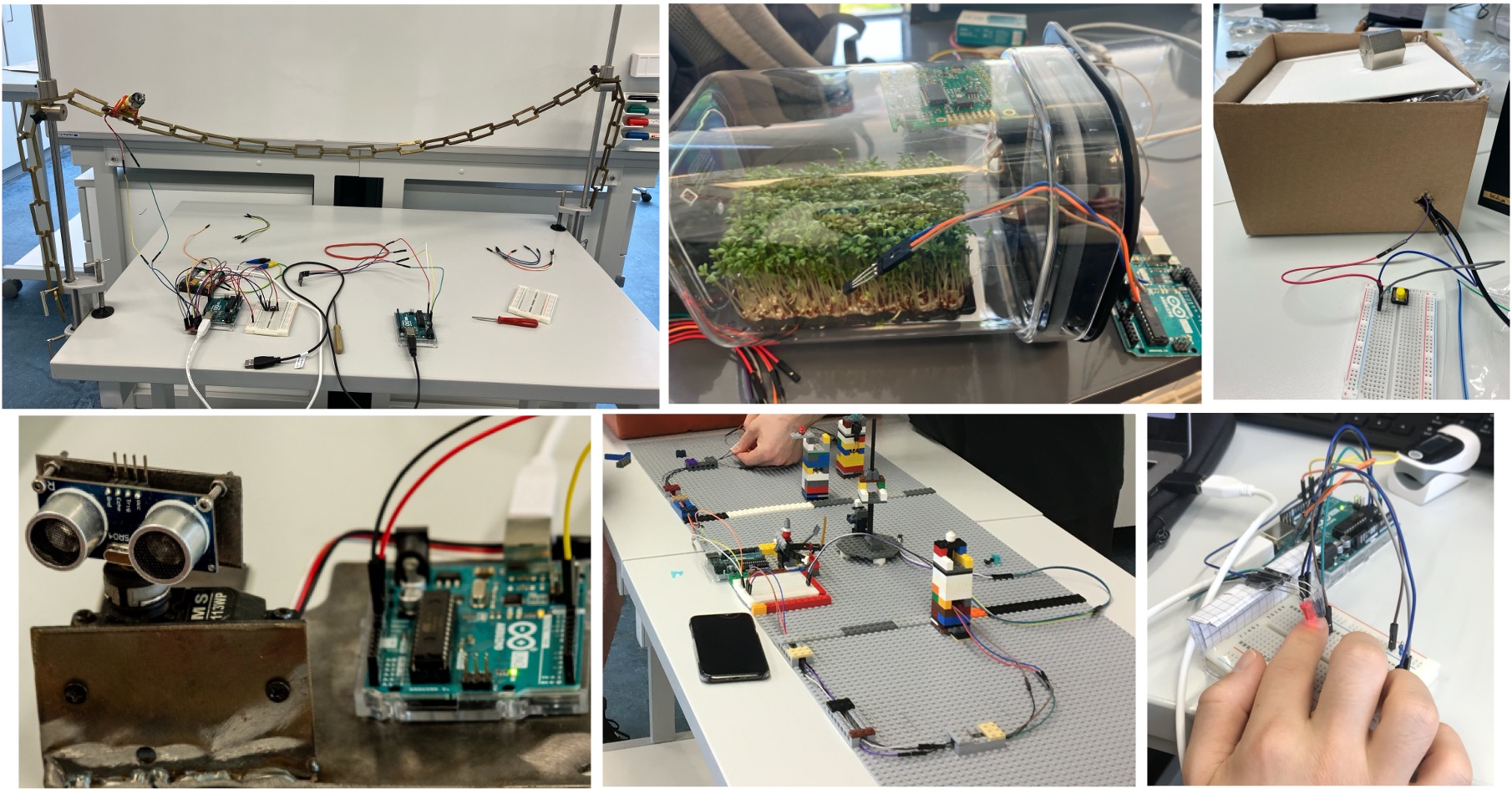}
\end{center}
\caption{\label{Projects} Example of setups designed and built by students during their projects.}
\end{figure}

An integral aspect of our scaffolding approach involves also the gradual reduction of explicit instructional guidance over time. While students receive substantial guidance through the use of probing questions at the onset of the course, this guidance gradually diminishes as students become proficient in several scientific skills and progress through the courses. Towards the end of the courses time, we installed students' projects, which foster greater independence and self-directed learning (an aspect discussed in the literature as \textit{agency} \cite{Agency}). In figure \ref{Projects} some examples of the setups designed and built by students during their projects are shown.

\subsection{Laboratory course structure}\label{CourseStructure}
As a result of the outlined scaffolding, we reorganized several components of the course structure. 

Prior to the transformation, various experiments with different setups were provided and students were randomly assigned to an order in which to process the experiments during the course of the semester.
While this approach allows for a reduced number of each setup and a large variety of experiments, it limits scaffolding for skill acquisition. Consequently, we eliminated the rotating structure and transitioned to a uniform activity structure for the majority of activities, ensuring all students engage in the same scaffolded tasks in the same order.

The scaffolding strategy outlined in the preceding section \ref{Scaffolding} has led to a decrease in the variety of experiments conducted by students and the range of physics topics covered (as detailed in the subsequent section \ref{PhysCont}). Presently, students do not switch experiments at every lab session but may continue working on the same experiment across multiple appointments.

\subsection{Physics Context}\label{PhysCont}
Before the transformation, the laboratory course covered many of the physics topics treated during the physics introductory lectures, spanning from mechanics and fluid dynamics to optics, thermodynamics, atomic physics and radioactivity.
Since the reconstructured lab course is not aimed to reinforce physics concepts studied in lecture, we identify three physics areas as particularly suitable for the acquisition of experimental skills:  Mechanics, Electricity/Electronic and Optics.
Mechanics was chosen for its simplicity in apparatus, electricity/electronics and optics were chosen because they allow students to practice troubleshooting and iteration \cite{AAPT1998, Zwickl3}.

\subsection{Laboratory space redesign}
We redesigned the laboratory space to foster a culture of collaboration and innovation among students. The introduction of open discussion areas facilitates group interactions during tutorials and experimental sessions, promoting peer-to-peer learning and knowledge exchange. Each room has been equipped with whiteboards, flip chart boards, and smart boards to encourage interactive learning. The integration of versatile furniture arrangements enables flexible use of space, accommodating varying group sizes and project requirements. Furthermore, the provision of fabrication tools, such as soldering stations, 3D printing facilities and very recently, a laser cutter, empowers students to engage in rapid prototyping and to materialize their design concepts effectively. 

\begin{figure}
\begin{center}
\includegraphics[width=1\textwidth]{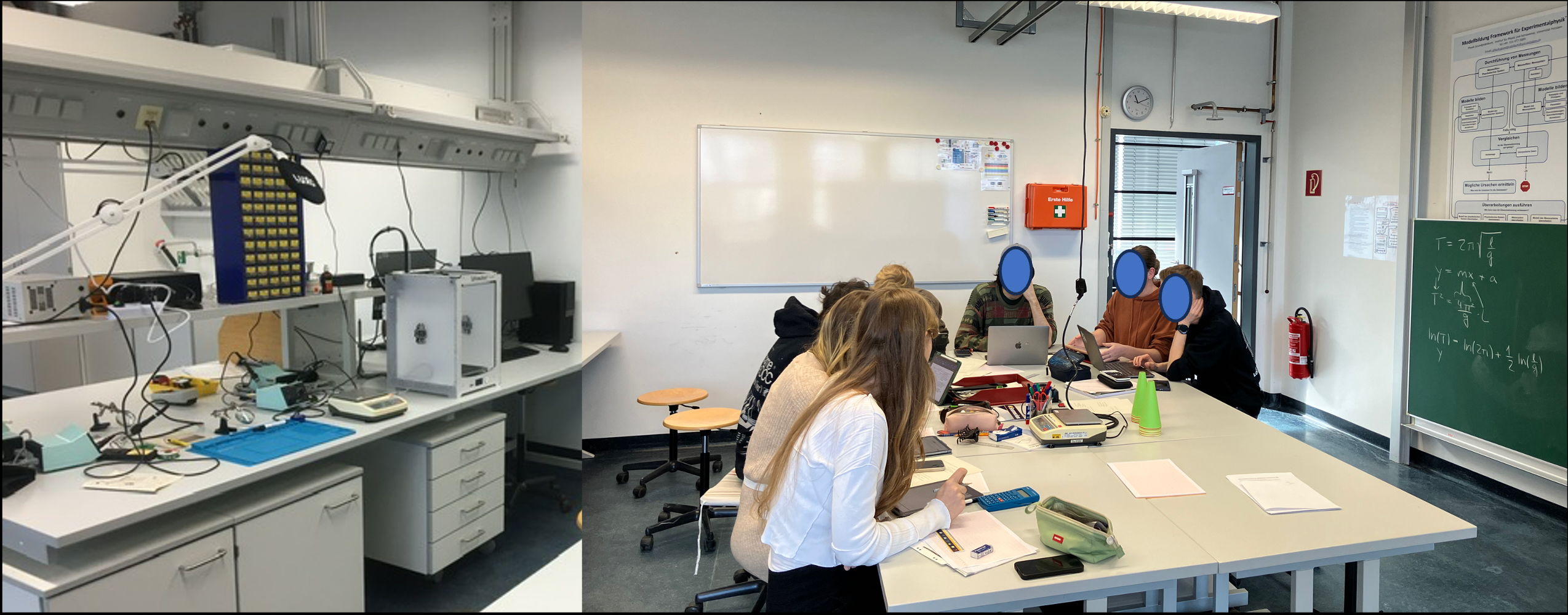}
\end{center}
\caption{\label{Labspace} Pictures of a lab area redesigned to allow students' discussions and foster communication.}
\end{figure}

The removal of pre-existing setups optimized space utilization, creating ample room for students' projects and promoting a conducive environment for creative exploration. We established a reading corner at the lab entrance, featuring journals and magazines to inspire experimental endeavors.
We attached the posters of the 'Modeling Framework for Experimental Physics' \cite{Zwickl, Zwickl3, Dounas} in each lab room. These posters are used by students and instructors while practicing modeling.
In figure \ref{Labspace}, one can see two views of an area for students discussion. One can notice a whiteboard, a blackboard, movable tables, and on the back the poster of the 'Modeling Framework for Experimental Physics' and a small sketchnote poster about lab notebooks. On the side (see figure on the left) there are soldering stations and a 3D printer. Not visible are on the other side a smartboard, a movable flipchart and benches for doing experimental work. Students in figure \ref{Labspace} are working together on solving data analysis problems.

\subsection{Laboratory Equipment}
Regarding the laboratory equipment, we prioritized leveraging existing resources while acquiring essential items aligned with our educational objectives. Notably, our learning outcomes did not hinge on the use of specialized or costly laboratory apparatus, as evidenced by the adaptability demonstrated during the COVID-19 pandemic, where students engaged in experimentation at home despite restricted access to expensive equipment. Nonetheless, our transformation necessitated the acquisition of both fundamental and sophisticated instruments. Notably, we transitioned predominantly to open-source software for data acquisition and analysis, minimizing the need for software investments. 

For the first semester, students utilize basic instruments from our inventory. For the second semester, we acquired electrical equipment and established four soldering stations to facilitate setting up customized electrical measurements (see figure \ref{Labspace}). Optical equipment such as breadboards, optics holders and different optical components was obtained for the third semester, enabling students to construct setups independently. 

For the fourth semester, we introduced Arduinos and associated materials, providing each student with a comprehensive kit including LEDs, resistors, cables, and sensors for in-class and at-home experimentation. This adaptation, initially prompted by the COVID-19 pandemic, has since become integrated into regular coursework with modifications for in-person instruction. The use of Arduinos reduces investment costs. Additional sensor or material purchases are made as needed for project work, averaging 100-200 Euros per semester. Students are encouraged to explore available resources and coordinate with instructors for equipment borrowing, ensuring safe and efficient project execution.

\subsection{Use of authentic forms of laboratory notebooks}
In the traditional class format prior to the redesign, students wrote comprehensive final reports detailing their experimental setups, procedures, data analysis, and conclusions following each lab session. This process demanded substantial time investment from students, overshadowing the documentation of in-lab activities within their reports.

To foster deeper engagement in scientific practices during laboratory sessions and to emphasize the importance of laboratory notebooks for professional experimental physicists, we opted to transition our assessment focus from laboratory reports to laboratory notebooks (LNs). As noted by Edward H. Carlson, "learning to keep a good lab notebook is a difficult skill and it is tightly coupled to good work habits" \cite{Carlson}.
We now require final reports only for students' projects, not for all the other activities in the course. Student performance is primarily evaluated based on the quality of their LNs for each lab activity. The pedagogical advantages of utilizing LNs have been previously underscored in the literature \cite{Stanley, Stanley2, Hoehn}.
They allow instructors to redirect attention from mere end-results to the entire experimental process. Instructors have the opportunity to recognize and commend students for acknowledging and reflecting on their mistakes within their lab notebooks. Additionally, LNs facilitate the documentation of iterative processes, such as model refinement and experimental apparatus adjustments, fostering the development of expert-like attitudes among students.
Furthermore, LNs serve as a platform to instill crucial habits, such as immediate data plotting during or directly after measurements, enabling students to closely track experimental progress and make informed decisions based on their findings. Encouraging students to reflect on their work within the lab setting empowers them to refine their experimental techniques and enhance their overall learning experience.

\subsection{Rubrics}
As recommended in the literature \cite{Etkina}, we initiated the implementation of rubrics for student assessment within our laboratory courses, recognizing immediately their benefits.
Primarily, rubrics serve to align student assessment with established learning objectives, thus mitigating the risk of divergent assessment among different instructors. Additionally, the process of crafting rubrics enables curriculum designers to scrutinize the validity of predetermined learning objectives, providing a tangible means to evaluate what is effectively practiced and measured within the curriculum.
Furthermore, rubrics empower students to focus their efforts on the key facets for which each activity was designed, offering them a structured guide to orient their work effectively.

\subsection{Students' coaching by instructors and teaching assistants}
To foster students' decision making and self-confidence in the laboratory  we reconsidered the role and the teaching practices of the laboratory instructors. 
They slipped into the role of facilitators that support students in taking decisions in a non-invasive way. They now use their experience while authentically discussing students' ideas in a critical way.
Enabling students to make decisions and attempt to resolve issues independently, as well as allowing them to practice troubleshooting, necessitates effective and coordinated interaction between instructors and students. 

To calibrate instructors interventions based on students' self-reflection about their experimental process status, we adopted a system from colleagues at the University of Hannover \cite{Cones}. Each group of students is provided with three cones: green, yellow, and red (see detail on the table of figure \ref{Labspace}). The green cone is exposed by students to indicate that they are working smoothly without encountering difficulties. When students encounter problems or troubleshooting scenarios, they display the yellow cone, and when they have persistently attempted to resolve a problem without success, they display the red cone. This visual cue allows instructors to tailor their interventions based on the students' expressed state, for example avoiding seeking interaction with students too often when they feel they are working smoothly on a process. 
There are occasions when students prefer to troubleshoot independently, even when encountering difficulties, demonstrating their desire to persevere and resolve issues.

\subsection{Group work}\label{GroupWork}
Before the laboratory transformation, students typically worked in pairs, a practice initially retained under the assumption that smaller groups would afford more hands-on practice and deeper learning experiences. However, insights gleaned from the Colorado Learning Attitudes about Science Survey for Experimental Physics (E-CLASS) survey \cite{Wilcox1}regarding students' attitudes towards communication prompted a reassessment of this approach \cite{Teichmann}. The findings revealed a lack of perceived importance regarding peer communication in scientific practices, prompting a strategic pivot in our instructional methods. To address this concern, we reconsidered how we foster communication in the entire lab course and added tasks or activities specifically focused on letting students brainstorm, discuss, present their ideas and results to other peers. 
For example, we introduced new activities wherein students collaborated in larger groups with shared goals \cite{Alemani_course}. Within these activities, students subdivided themselves into smaller teams, each tasked with specific sub-goals, fostering intergroup communication to exchange insights, solicit feedback, and resolve challenges collectively. An example of such a collective task is to build and test a radio receiver in the middle waves.

Moreover, tutorial sessions were restructured to accommodate larger group sizes, ranging from 3 to 12 students depending on the nature of the tasks. Notably, we adjusted the typical group size for experiments to three individuals and up to four for projects. This modification facilitates real-time experimentation, data analysis (utilizing Python), and documentation through shared responsibilities within the enlarged groups. Emphasizing the importance of structured group dynamics, we implemented mechanisms to ensure equitable participation, such as rotating roles during sessions or assigning a student as a communication facilitator to streamline intra-group exchanges or we reserved time for students' presentations to peers.

\subsection{Description of the curriculum}
The curriculum resulting from the above described transformation process is now coarsely described. A description of the scaffolded activities can be found in \cite{OSF}.

In the first semester, students first learn the fundamentals of measurement uncertainties, data analysis, Python programming and scientific documentation during active learning tutorials. Thereafter, they practice those skills further during laboratory activities. 
In the second semester we introduce \textit{modeling} using the \textit{Modeling Framework for Experimental Physics} \cite{Zwickl, Zwickl3, Dounas} and tailored activities. Students practice modeling electrical circuits and apparatus, identifying the assumptions and limits of those models and refining them. For example students have to figure out electrical components inside a closed box. At the end of the semester students participate in a large group project in which they practice more independently working with electrical circuits.

Students continue practicing on \textit{modeling} and the skills in the other categories in the third semester by designing, building, and characterizing various optical measurement apparatus.

In the fourth semester, students extend their technical skills learning how to use and programm Arduino microprocessors and to measure or modify the real world by means of sensors and actuators. In different experiments students continue practicing modeling. At the end of the semester, during projects, they work on designing experiments to answer a self-determined research question (see figure  \ref{Projects}).

\section{Assessment}
We used the E-CLASS survey to assess and guide our course transformation. The results until 2022 can be found in references \cite{Teichmann, Teichmann2}. Here we just summarize the main findings. Instruction had a positive impact on students views about experimental physics. In particular, students gain confidence in doing data analysis, in overcoming difficulties and in using new lab equipment. On the other hand, instruction had a negative impact regarding the importance of communicating scientific results to peers. As mentioned previously in section \ref{GroupWork}, this result motivated us to focus on communication aspects in the years after the assessment. To monitor the effects of our changes regarding this aspect we plan to collect further E-CLASS data.

\section{Conclusions}
Our article has outlined the comprehensive redesign and reconstruction of an introductory laboratory course for physics major students. The goal is to engage students actively and provide them with genuine experiences in scientific practices while cultivating their acquisition of experimental skills and expert-like attitudes. We have delineated the newly defined learning objectives and the changes imparted to the existing course and the activities crafted to activate students and immerse them in high-level cognitive tasks. 

This work serves as a valuable resource for educators embarking on similar endeavors, offering practical guidance and innovative strategies to enhance student engagement and promote active learning in laboratory settings.

\ack{}
The author would like to thank all who contribuited or supported the laboratory reconstruction process. Great gratitude goes to all students who provided their precious and continuous feedback and consistently encourcourage the author in her efforts. I acknowledge support from the Stifterverband and the Baden-Württemberg Stiftung (Project No. H1205228500830925).

\end{document}